\documentclass[aip,jcp,reprint,preprintnumbers,amsmath,amssymb,floatfix,nofootinbib]{revtex4-1}

\usepackage{graphicx}
\usepackage{color}
\usepackage{amsmath}
\usepackage{mathtools}
\usepackage{amssymb}
\usepackage{amsfonts}  
\usepackage{rotating}
\usepackage{multirow}
\usepackage{dcolumn}
\usepackage{bbold}

\begin{document}
\edef\marginnotetextwidth{\the\textwidth}

\title{A real-time extension of density matrix embedding theory for non-equilibrium electron dynamics} 

\author{Joshua S. Kretchmer}
\author{Garnet Kin-Lic Chan}
\email{gkc1000@gmail.com}
\affiliation{Divsion of Chemistry and Chemical Engineering, California Institute of Technology, Pasadena, California 91125, USA}

\date{\today}

\begin{abstract}
We introduce real-time density matrix embedding theory (DMET), a dynamical quantum embedding theory for computing non-equilibrium electron dynamics in strongly correlated systems.
As in the previously developed static DMET, real-time DMET partitions the system into an impurity corresponding to the region of interest coupled to the surrounding environment, which is efficiently represented by a quantum bath of the same size as the impurity.
In this work, we focus on a single-impurity time-dependent theory as a first step towards a full multi-impurity theory. 
The equations of motion of the coupled impurity and bath embedding problem in real-time DMET are then derived using the time-dependent variational principle.
The accuracy of real-time DMET is compared to that of time-dependent complete active space self-consistent field (TD-CASSCF) theory and time-dependent Hartree-Fock (TDHF) theory for a variety of quantum quenches in the single impurity Anderson model (SIAM), in which the Hamiltonian is suddenly changed (quenched) to induce a non-equilibrium state. Real-time DMET shows a marked improvement over the meanfield TDHF, converging to the exact answer even in the non-trivial Kondo regime of the SIAM. However, as in static DMET, the increased dynamic flexibility of TD-CASSCF in comparison to real-time DMET using a single impurity leads to faster convergence with respect to active space size.
Our results demonstrate that real-time DMET is an efficient method well suited for the simulation of non-equilibrium electron dynamics in which strong electron correlation plays an important role.
\end{abstract}

\maketitle 

\section{Introduction}

Non-equilibrium electron dynamics is prevalent throughout chemistry and physics, for example, in electron 
transport through molecular junctions,\cite{Avi74,Joa00,Nit03,Hea03} electron injection and transport following photoexcitation,\cite{Hag10,Kis14} and driven electron dynamics in laser pulses.\cite{Kra09,Gal12} The simulation of such processes is challenging due to the need to treat both large system sizes and electron correlation. 
A variety of methods have been developed for non-equilibrium electron dynamics, including non-equilibrium Green's function approaches,\cite{Aok14,Fre06,Ram98,Mit04,Gal06,Ryn06,Har08,Tah08} numerical path-integral techniques,\cite{Muh08,Wei08,Wei09,Seg10,Eck10} real-time Monte Carlo methods,\cite{Coh15,Wer09,Wer06,Sch10,Gul11,Gul11b,Coh14,Coh14b} semiclassical approximations,\cite{Li13,Swe11,Swe12} and wavefunction propagation methods.\cite{Wan13,Wan09,Lin15,Zan03,Alo07,Kat04,Sat13,Mir11,Nes05,Roh06,Gre10,Run84,Yab96,McL64,Kul87,Rin80,Caz02,Luo03,Sch04,Whi04,Dal04} In this work,
we will present new developments in the latter class.

Most wavefunction-based methods fall into two categories: those which sacrifice accuracy for the ability to treat large system sizes, such as time-dependent density functional theory\cite{Run84,Yab96,Ale14,Pen15,Mak15,Mak16} and time-dependent Hartree-Fock theory,\cite{McL64,Kul87,Rin80} and methods which are highly accurate, but are limited to small system sizes, such as multi-configurational time-dependent Hartree theory\cite{Wan13,Wan09} and the time-dependent density matrix renormalization group (DMRG).\cite{Caz02,Luo03,Sch04,Whi04,Dal04,Alhas06,Hei09} To improve
the compromise between accuracy and efficiency, we here borrow an idea from electronic structure approximations, namely
that of quantum embedding. Embedding techniques work by dividing the total system into a small region of interest, termed the impurity, which is treated accurately, and the surrounding environment, which is treated in an approximate manner. This decomposition allows calculations
on a large total system, while retaining a high-level of accuracy in the region of interest.

One powerful embedding formulation that has  been introduced for static electronic properties
 is the density matrix embedding theory (DMET).\cite{Kni12,Kni13,Wou16} In DMET, the surrounding environment is represented by a quantum bath,  constructed to capture the entanglement between the environment and the impurity. The entanglement-based construction ensures that the size of the quantum bath is at most equal to the size of the impurity. The bath allows
for strong coupling between the impurity and environment,
while its small size ensures computational efficiency. 
Furthermore, in a general DMET calculation, the total system can be divided into multiple local impurities, each associated with its own embedded problem.
DMET has been successfully applied to fermion and spin lattice models\cite{Kni12,Zhe16,Boo15,Che14,Bul14,Zhe16b} as well as ab initio molecular and condensed phase systems.\cite{Kni13,Wou16,Bul14b,Tsu15}

Here, we will use the advantages of a quantum embedding formulation
for dynamics, by extending  DMET to the real-time propagation of the electronic wavefunction. 
Specifically, we focus on a real-time extension of the single impurity formulation of DMET as a first step in the possible development of a full multi-impurity time-dependent theory.
In this case, 
the quantum bath  becomes time-dependent. We use the time-dependent variational principle (TDVP)\cite{Fre34,Low72,Moc73} to derive the dynamics of the quantum bath,
as well as that of the correlated wavefunction in the coupled impurity-bath problem. 
However, we introduce a constraint in the TDVP that the impurity orbitals remain time-independent, which is necessary to maintain the definition of the impurity during the dynamics.
The real-time DMET possesses analogous formal strengths 
to the original static formulation,
and provides an exact description of dynamics in the non-interacting, isolated cluster, and large impurity size limits. 
We demonstrate the
strengths of the method by simulating several kinds of quantum quenches in the single impurity Anderson model (SIAM),
comparing, where possible, against time-dependent Hartree Fock (TDHF) theory, time-dependent complete active space self-consistent field (TD-CASSCF) theory, and numerically exact time-dependent density matrix renormalization group (TD-DMRG) benchmarks.\cite{itensor,Alhas06} 
A quantum quench is defined by a sudden change in the Hamiltonian, which induces a non-equilibrium state in the system and thus subsequent dynamics.
We find excellent numerical agreement between real-time DMET and the numerically exact TD-DMRG, including in the regime of the Kondo resonance.
Overall, our results show that DMET offers an accurate treatment of the quantum dynamics in the impurity region,
with a very affordable cost.

\section{Theory}\label{sec:theory}

\subsection{Static DMET}\label{sec:statdmet}

We begin by reviewing the static DMET algorithm to provide a foundation for the later presentation of real-time DMET. 
For simplicity, we will assume that the static problem of interest is the ground-state problem, and will focus
only on the ``interacting bath'' formulation of DMET.\cite{Wou16} 
We also restrict our discussion to a single embedded impurity cluster, as described below.

Consider a full quantum system spanned by an orthonormal single-particle basis, indexed by $p$, $q$, $r$ and $s$. 
 The starting, local, single-particle basis of the problem will be referred to as {\it sites}, while more
general single-particle functions  will be termed orbitals. The total size of the basis will be denoted $N$. The general second-quantized Hamiltonian for the 
full system can be written as
\begin{equation}
\hat{H}=\sum_{pq}h_{pq}E_{pq}+\frac{1}{2}\sum_{pqrs}V_{pqrs}E_{pqrs},\label{eqn:ham_gen}
\end{equation}
where $h_{pq}=\langle p|\hat{h}|q\rangle$ and $V_{pqrs}=\langle pq|\hat{V}|rs\rangle$ are the one-and two-electron Hamiltonian matrix elements,
\begin{equation}
E_{pq}=\sum_{\sigma}a^\dag_{p\sigma}a_{q\sigma}
\end{equation}
and
\begin{equation}
E_{pqrs}=\sum_{\sigma\tau}a^\dag_{p\sigma}a^\dag_{q\tau}a_{s\tau}a_{r\sigma}.
\end{equation}
The operator $a^\dag_{p\sigma}$ ($a_{p\sigma}$) creates (destroys) an electron of spin $\sigma$ at site $p$. 

The single particle basis of the full quantum system can be partitioned into a small subset of sites, $i\in A$, corresponding to the region of interest and termed the impurity; the number of these sites will be denoted $N_\text{imp}$. The remainder of the sites constitute the environment. Static DMET relies on the observation that,
for any state of the full quantum system, 
 the entanglement between the impurity and the surrounding environment can be exactly accounted for by a quantum bath that is the same size as the impurity. Specifically, given the exact ground-state of the full quantum system, it 
can be written through its Schmidt decomposition as 
\begin{equation}
|\Psi\rangle=\sum_{i}^{M_A}
\psi_{i}|\alpha_i\rangle|\beta_i\rangle,
\end{equation}
where, $\psi_{i}$ is an expansion coefficient, $|\alpha_i\rangle$ are (multi-electron) states in the Fock space spanned by the impurity $A$, and $|\beta_i\rangle$ are (multi-electron) states in the Fock space of the environment that constitute the quantum bath. Note that though the states $|\beta_i\rangle$ fully capture the entanglement with the environment, there are only $M_A$ of them: the dimension of the Fock space of the impurity. 
Thus, in principle, the ground-state can be determined by solving the Schr\"{o}dinger equation with $\hat{H}$ projected into the 
small impurity plus bath Hilbert space. 
However, this is not a practical solution, as
the definition of the environment states $|\beta_i\rangle$ requires knowledge of the exact solution. To circumvent 
this, in static DMET the states $|\beta_i\rangle$ are calculated instead from the ground-state of a simpler Hamiltonian, $\hat{h}'$. 
The static DMET approximations to the  ground-state and expectation values of the original interacting problem 
thus require self-consistently solving two coupled models: (i) for the ground-state, $|\Phi\rangle$, of the approximate Hamiltonian, $\hat{h}'$, in the full system Hilbert space and (ii) for the ground-state, $|\Psi_{\mathrm{imp}}\rangle$, of the interacting problem, within the
{\it small} embedding Hilbert space of the impurity coupled to the now approximate quantum bath.
 A self-consistency condition on the one-particle reduced density matrix links the two models.

In most applications of static DMET, the approximate Hamiltonian for the full quantum system is defined as a single-particle Hamiltonian of the form
\begin{equation}
\hat{h}'=\hat{h}+\hat{u},
\end{equation}
where, $\hat{h}$ is  most commonly chosen to be either the one-particle part of the total Hamiltonian, $\hat{H}$, or the Fock operator derived from $\hat{H}$. In the case of purely local interactions, as in the Anderson impurity model studied in this work, the two choices are equivalent. Here, $\hat{u}$ is the local {\it correlation potential},
\begin{equation}\label{eqn:uhat}
\hat{u}=\sum_{pq\in A}u_{pq}E_{pq},
\end{equation}
which approximates the effect of the local Coulomb interaction within the impurity. The sum in Eq. \eqref{eqn:uhat} is restricted to the $N_{\mathrm{imp}}$ sites within the impurity $A$. The elements $u_{pq}$ are obtained through the self-consistency condition described below.

The quantum bath that defines the embedding problem is obtained from the ground-state of $\hat{h}'$, $|\Phi\rangle$, which takes the form of a simple Slater determinant. In this case, the multi-electron states, $|\beta_i\rangle$, assume a particularly simple form: they constitute states in the Fock space spanned by a set of at most $N_{\mathrm{imp}}$  bath orbitals, multiplied by a core determinant. These  {\it embedding} orbitals can be obtained in several mathematically equivalent ways. Here, we will assume that the bath orbitals are determined
by diagonalizing part of the one-particle density matrix, $\rho^{\Phi}$, computed from $|\Phi\rangle$, with elements $\rho^{\Phi}_{pq}=\langle \Phi|E_{qp}|\Phi\rangle$. The one-particle density matrix can be partitioned into a $N_{\mathrm{imp}}\times N_{\mathrm{imp}}$ impurity block, a $(N-N_{\mathrm{imp}})\times(N-N_{\mathrm{imp}})$ environment block, and $N_{\mathrm{imp}}\times(N-N_{\mathrm{imp}})$ off-diagonal coupling blocks,
\begin{equation}\label{eqn:rhomf}
\rho_\Phi^{\sigma}\equiv
\begin{bmatrix}
    \rho_{\text{imp}}^\Phi&\rho_{\text{c}}^\Phi\\
    \rho_{\text{c}}^{\Phi\dagger} & \rho_{\text{env}}^\Phi
    \end{bmatrix}.
\end{equation}
Diagonalizing the environment block of the one-particle density
matrix, $\rho_{\mathrm{env}}=R_{\mathrm{env}}\Lambda
R_{\mathrm{env}}^\dag$, yields three kinds of embedding orbitals: (i) a
set of $N_{\mathrm{imp}}$ {\it bath} embedding orbitals with eigenvalues
between zero and two, that describe entanglement between the environment and impurity,
(ii) a set of $N_{\mathrm{occ}}-N_{\mathrm{imp}}$ {\it core} embedding
orbitals, where $N_{\mathrm{occ}}$ is the number of occupied orbitals in
$|\Phi\rangle$, with eigenvalues equal to two, that are thus not
entangled with the impurity; these orbitals comprise the core
determinant, and (iii) a set of $N-N_{\textrm{imp}}-N_{\mathrm{occ}}$
{\it virtual} embedding orbitals, with eigenvalues equal to zero, that are thus also not entangled
with the impurity. The embedding problem thus consists of a complete
active space (CAS) wavefunction calculation in which the impurity and bath orbitals
comprise the active space, and the core determinant is comprised of
the core embedding orbitals; the virtual embedding orbitals constitute the
space external to the active and core spaces.

Before continuing, we introduce some notation for the embedding orbitals:  impurity orbitals will be designated by indices $i$ and $j$; 
 embedding bath orbitals will be designated by $y$ and $z$; the combination of all active space orbitals, which correspond to both the impurity and bath orbitals, will be designated by $l$, $k$, $m$ and $n$;  embedding core orbitals will be designated by $u$, $v$, $w$, and $x$;  embedding virtual orbitals will be designated by indices $a$ and $b$; the combination of all single-particle orbitals in the embedding basis will be designated by $c$, $d$, $e$, $f$, and $g$.

In the interacting-bath formulation of static DMET, the Hamiltonian of the embedding problem, $\hat{H}_{\mathrm{imp}}$, is obtained by projecting the original fully-interacting Hamiltonian, $\hat{H}$, into the active-space defined in the embedding basis, and including the contribution from the doubly occupied core determinant. This can be performed by a change of single-particle basis from the original site-basis to the embedding basis while including a contribution from the core orbitals, such that
\begin{equation}
\hat{H}_{\mathrm{imp}}=\sum_{lk}\tilde{h}_{lk}E_{lk}+\frac{1}{2}\sum_{lkmn}V_{lkmn}E_{lknm},\label{eqn:ham_emb}
\end{equation}
where 
\begin{eqnarray}
\tilde{h}_{cd}&=&h_{cd}+\sum_{u}\left(2V_{cudu}-V_{cuud}\right),\\
h_{cd}&=&\sum_{pq}R_{pc}^*h_{pq}R_{qd},
\end{eqnarray}
and
\begin{equation}
V_{cdef}=\sum_{pqrs}R_{pc}^*R_{qd}^*V_{pqrs}R_{re}R_{sf}.
\end{equation}
The rotation matrix from the site-basis to the embedding basis is given by
\begin{equation}\label{eqn:rotmat}
R=
\begin{bmatrix}
    \mathbb{1}_{N_{\mathrm{imp}}\times N_{\mathrm{imp}}}&0\\
    0& R_{\mathrm{env}}
    \end{bmatrix},
\end{equation}
where the identity matrix denotes that the impurity orbitals are the same in the original site basis and the embedding basis; $R_{\mathrm{env}}$ is defined above.

A wide range of solvers can be used to compute the correlated ground-state, $|\Psi_{\mathrm{imp}}\rangle$, of the embedding Hamiltonian, $\hat{H}_{\mathrm{imp}}$, depending on the nature of the problem as well as the cost and accuracy requirements. 
In this work we use exact diagonalization as the impurity solver, though previous work has also employed DMRG,\cite{Che14,Bul14,Zhe16b} coupled cluster theory,\cite{Wou16,Bul14} and auxiliary-field quantum Monte Carlo.\cite{Zhe16}

As described above, the elements of the correlation potential $\hat{u}$ are determined by a self-consistent procedure. Specifically, we minimize the difference between the impurity block of the one-body density matrices calculated from the uncorrelated wavefunction, $|\Phi\rangle$, and correlated wavefunction, $|\Psi_{\mathrm{imp}}\rangle$,
\begin{equation}
\min_u f(u) \ \text{where} \ f(u)=\sqrt{\sum_{ij\sigma}\left\{\rho^{\Phi}_{ij}-\rho^{\Psi_{\textrm{imp}}}_{ij}\right\}^2},\label{eqn:cost}
\end{equation}
and the elements $\rho^{\Psi_{\mathrm{imp}}}_{ij}=\langle \Psi_{\mathrm{imp}}|E_{ji}|\Psi_{\mathrm{imp}}\rangle$. However, as in previous work,\cite{Wou16,Zhe16,Kni13} the functional $f(u)$ is not directly optimized, but instead a self-consistent iteration is used: $f(u)$ is optimized with a fixed $|\Psi_{\mathrm{imp}}\rangle$; the optimal $u$ is then used to update $|\Phi\rangle$, the embedding Hamiltonian, $\hat{H}_{\mathrm{imp}}$, and thus $|\Psi_{\mathrm{imp}}\rangle$.

In summary, the static DMET algorithm proceeds via the following steps:
\begin{enumerate}
  \item we choose an initial guess for the correlation potential $\hat{u}$;\label{item:step1}
  \item we solve for the approximate Hamiltonian, $\hat{h}'$, to obtain the reference wavefunction $|\Phi\rangle$;
  \item we construct the embedding Hamiltonian using Eq. \eqref{eqn:ham_emb};
  \item we use exact diagonalization to compute the ground state of the embedding problem, $|\Psi_{\mathrm{imp}}\rangle$, and construct the one-body density matrix $\rho^{\Psi_{\mathrm{imp}}}$;
  \item we minimize $f(u)$ in Eq. \eqref{eqn:cost}, with $\rho^{\Psi_{\mathrm{imp}}}$ fixed, to obtain a new correlation potential $u^\prime$;
  \item if $||u-u^\prime||_\infty > \varepsilon_0$, the convergence threshold, we set $u=u^\prime$ and go to step~\ref{item:step1}; otherwise the static DMET calculation is converged.
\end{enumerate}
%

\subsection{Real-time DMET}\label{sec:rtdmet}

We now describe the central methodological contribution of the paper, namely, a real-time extension of  DMET, which 
allows for the efficient time-propagation of the electronic wavefunction. We focus on the propagation of the only a single impurity and its corresponding quantum bath as a first step in the development of a multi-impurity time-dependent framework.

Developing a real-time extension of DMET entails discerning how to appropriately propagate (i) the embedding
orbitals, to give the approximate representation of the time-dependent environment, and (ii) the correlated
CAS-like DMET wavefunction in the embedding problem, to describe the region of interest at a high-level.
Here, we utilize the TDVP to derive the equations of motion for both the embedding orbitals and the expansion coefficients for the determinants in the DMET CAS-like wavefunction. 
We introduce the constraint that the impurity orbitals are time-independent in keeping
with an embedding picture in which one can cleanly identify the contributions of different impurities to global observables: even though we only consider a single impurity formulation here, we keep this picture in anticipation of the future development of the multiple impurity formalism.
Our TDVP derivation intrinsically connects the low-level orbital dynamics and
the high-level embedded dynamics, and thus does not require a further self-consistency through a
time-dependent correlation potential.
 We return to the question of the challenges of a self-consistent picture from a time-dependent correlation potential in App. \ref{app:vrep}.
We now present the detailed derivation of the equations of motion, which constitute the working equations for the real-time DMET method.

To begin, we write the correlated wavefunction for the embedding problem as a time-dependent CAS wavefunction,
\begin{equation}
|\Psi_{\mathrm{imp}}(t)\rangle=\sum_{M}C_M(t)|M(t)\rangle,\label{eqn:psi_imp}
\end{equation}
where $|M(t)\rangle$ are time-dependent determinants in the active space defined by the impurity and bath orbitals coupled to a doubly-occupied determinant comprised of the embedding core orbitals, and $C_M(t)$ are the time-dependent expansion coefficients. The time-dependence of the determinants arises from the time-dependence of the embedding bath and core orbitals, as the impurity orbitals are kept time-independent.

Following the TDVP,\cite{Fre34,Low72,Moc73,Mir11,Sat13} the equations of motion for both $|M(t)\rangle$ and $C_M(t)$ can be obtained by varying the Dirac-Frenkel action with fixed endpoints,
\begin{equation}
S[\Psi_{\mathrm{imp}}]=\int_{t_0}^{t_1}dt\langle\Psi_{\mathrm{imp}}|\hat{H}-i\hbar\frac{\partial}{\partial t}|\Psi_{\mathrm{imp}}\rangle.
\end{equation}
This procedure yields the variational equation
\begin{eqnarray}
\left\langle \delta\Psi_{\mathrm{imp}}|\left(\hat{H}-i\hbar\frac{\partial}{\partial t}\right)\Psi_{\mathrm{imp}}\right\rangle+\nonumber\\\left\langle \left(\hat{H}-i\hbar\frac{\partial}{\partial t}\right)\Psi_{\mathrm{imp}}|\delta\Psi_{\mathrm{imp}}\right\rangle=0,\label{eqn:var}
\end{eqnarray}
which must be satisfied for arbitrary variations of the wavefunction, $\delta\Psi_{\mathrm{imp}}$. We should note that the ground-state DMET wavefunction is not rigorously a stationary state of the TDVP since the ground-state DMET is not variationally optimized. However, numerical investigation seems to suggest that the difference is small and should not provide any issues in the future use of the methodology.

The variation of the wavefunction with respect to the expansion coefficients can be written as
\begin{equation}
|\delta_C\Psi_{\mathrm{imp}}\rangle=\sum_M\delta C_M|M\rangle,\label{eqn:var_coef}
\end{equation}
while the variation with respect to the embedding  orbitals can be written as\cite{Mir11}
\begin{equation}
|\delta_a\Psi_{\mathrm{imp}}\rangle=\sum_{ab}\Delta_{ab}E_{ab}|\Psi_{\mathrm{imp}}\rangle,\label{eqn:var_orb}
\end{equation}
where $\Delta_{ab}$ is an anti-Hermitian matrix. The complete time-dependence of the wavefunction can be expressed as
\begin{eqnarray}
i\hbar|\dot{\Psi}_{\mathrm{imp}}\rangle&=&i\hbar\sum_M\dot{C}_M|M\rangle+C_M|\dot{M}\rangle\\
&=&\sum_Mi\hbar\dot{C}_M|M\rangle+C_M\hat{X}|M\rangle,\label{eqn:dotpsi}
\end{eqnarray}
where we have introduced the single-particle Hermitian operator $\hat{X}$, which governs the time-dependence of the embedding orbitals. The operator is defined as 
\begin{equation}
\hat{X}=\sum_{cd}X_{cd}E_{cd},
\end{equation}
where the elements $X_{cd}=i\hbar\langle c|\dot{d}\rangle$ are determined through the variational equation, Eq. \eqref{eqn:var}, as shown below.

Inserting the variation with respect to the expansion coefficients, Eq. \eqref{eqn:var_coef}, and the time-dependence of the wavefunction, Eq. \eqref{eqn:dotpsi}, into the variational equation, Eq. \eqref{eqn:var}, yields
\begin{equation}
i\hbar\dot{C}_M=\sum_N\langle M|\left(\hat{H}-\hat{X}\right)|N\rangle C_N,\label{eqn:dotc}
\end{equation}
which defines the equations of motion of the expansion coefficients.
Inserting the variation with respect to the  orbitals, Eq. (\ref{eqn:var_orb}), yields
\begin{eqnarray}
\langle\Psi_{\mathrm{imp}}|\left(\hat{H}-\hat{X}\right)\left(1-\Pi\right)E_{ab}|\Psi_{\mathrm{imp}}\rangle\nonumber\\
-\langle\Psi_{\mathrm{imp}}|E_{ab}\left(1-\Pi\right)\left(\hat{H}-\hat{X}\right)|\Psi_{\mathrm{imp}}\rangle=0,\label{eqn:orb1}
\end{eqnarray}
where $\Pi=\sum_M|M\rangle\langle M|$ is the projector into the CAS space defined by the impurity and embedding orbitals. Solving Eq. \eqref{eqn:orb1} 
defines the elements of the operator $\hat{X}$.

The elements of $\hat{X}$ will now be derived for each type of orbital rotation in the embedding basis, utilizing the notation for the 
different kinds of embedding orbitals defined in Sec. \ref{sec:statdmet}.
Eq. \eqref{eqn:orb1} reduces to a trivial identity for an orbital pair $\{c,d\}$ corresponding to the same orbital subspace (core, active, or virtual) due to the presence of the projector $(1-\Pi)$ out of the CAS space;\cite{Mir11,Sat13} the determinants $\hat{E}_{uv}|M\rangle=2\delta_{uv}|M\rangle$, $\hat{E}_{lk}|M\rangle$, and $\hat{E}_{ab}|M\rangle=0$ are either zero or fall within the CAS space and are thus eliminated by the projector $(1-\Pi)$. These intraspace orbital rotations, $\hat{E}_{cd}=\{\hat{E}_{uv},\hat{E}_{lk},\hat{E}_{ab}\}$ are referred to as redundant, since the total wavefunction is invariant to such rotations if accompanied by the corresponding transformation of the expansion coefficients as seen by the presence of $\hat{X}$ in Eq. \eqref{eqn:dotc}.\cite{Mir11,Sat13,Cai05,Hel13} The elements of $\hat{X}$ for these redundant orbital pairs can then be freely chosen; in this work we set these terms to zero such that $X_{uv}=X_{vu}^*=X_{lk}=X_{kl}^*=X_{ab}=X_{ba}^*=0$.

For the non-redundant orbital pairs, the projector, $(1-\Pi)$, in Eq. \eqref{eqn:orb1} can be dropped and the equation reduces to
\begin{eqnarray}
\sum_e\left[X_{ce}\rho^{\Psi_{\mathrm{imp}}}_{ed}\right.&-&\left.\rho^{\Psi_{\mathrm{imp}}}_{ce}X_{ed}\right]=\sum_e\left[h_{ce}\rho^{\Psi_{\mathrm{imp}}}_{ed}-\rho^{\Psi_{\mathrm{imp}}}_{ce}h_{ed}\right]\nonumber\\
&+&\sum_{def}\left[V_{ecgf}\Gamma^{\Psi_{\mathrm{imp}}}_{fgde}-V_{efgd}\Gamma_{gcef}^{\Psi_{\mathrm{imp}}}\right],\label{eqn:orb2}
\end{eqnarray}
where the 2-particle reduced density matrix has elements $\Gamma_{cdef}^{\Psi_{\mathrm{imp}}}=\langle\Psi^{\mathrm{imp}}|E_{efcd}|\Psi^{\mathrm{imp}}\rangle$. Eq. \eqref{eqn:orb2} can now be solved for each non-redundant orbital rotation. 

As mentioned above, the impurity orbitals in real-time DMET are restricted to be time-independent. Therefore, all elements of $\hat{X}$ that include an impurity orbital are defined to be zero, such that $X_{ic}=X_{ci}^*=0$. 

The other non-redundant orbital rotations can be obtained using the non-zero elements of the reduced density matrices for the embedding wavefunction, which are $\rho^{\Psi_{\mathrm{imp}}}_{uv}=2\delta_{uv}$, $\rho^{\Psi_{\mathrm{imp}}}_{lk}$, $\Gamma^{\Psi_{\mathrm{imp}}}_{uvwx}=4\delta_{uw}\delta_{vx}-2\delta_{ux}\delta_{vw}$,  $\Gamma^{\Psi_{\mathrm{imp}}}_{luku}=\Gamma^{\Psi_{\mathrm{imp}}}_{uluk}=2\rho^{\Psi_{\mathrm{imp}}}_{lk}$, $\Gamma^{\Psi_{\mathrm{imp}}}_{luuk}=\Gamma^{\Psi_{\mathrm{imp}}}_{ulku}=-\rho^{\Psi_{\mathrm{imp}}}_{lk}$, and $\Gamma^{\Psi_{\mathrm{imp}}}_{lkmn}$. This then yields
\begin{eqnarray}
X_{au}=X_{ua}^*&=&\tilde{h}_{au}+\sum_{lk}\left(V_{kalu}-\frac{1}{2}V_{kaul}\right)\rho^{\Psi_{\mathrm{imp}}}_{lk},\label{eqn:xcorevirt}\\\nonumber\\
X_{az}=X_{za}^*&=&\sum_y\left[\sum_{k}\tilde{h}_{ak}\rho^{\Psi_{\mathrm{imp}}}_{ky}+\sum_{lkm}V_{lamk}\Gamma^{\Psi_{\mathrm{imp}}}_{kmyl}\right]\nonumber\\
&&\left[(\rho^{\Psi_{\mathrm{imp}}}_{\mathrm{bath}})^{-1}\right]_{yz},\label{eqn:xbathvirt}
\end{eqnarray}
and
\begin{eqnarray}
X_{zu}=X_{uz}^*&=&\sum_y\left[(\bar{\rho}^{\Psi_{\mathrm{imp}}}_{\mathrm{bath}})^{-1}\right]_{zy}\left(2\tilde{h}_{yu}-\sum_k\rho^{\Psi_{\mathrm{imp}}}_{yk}\tilde{h}_{ku}\right.\nonumber\\
&+&\left.\sum_{kl}\left(V_{lyku}-V_{lyuk}\right)\rho^{\Psi_{\mathrm{imp}}}_{kl}\right.\nonumber\\
&-&\left.\sum_{klm}V_{klmu}\Gamma^{\Psi_{\mathrm{imp}}}_{mykl}\right),\label{eqn:xbathcore}
\end{eqnarray}
where the matrix $\rho^{\Psi_{\mathrm{imp}}}_{\mathrm{bath}}$ corresponds to the bath block of the 1-electron reduced density matrix and $\left[\bar{\rho}^{\Psi_{\mathrm{imp}}}_{\mathrm{bath}}\right]_{yz}=2\delta_{yz}-\rho^{\Psi_{\mathrm{imp}}}_{yz}$.

The real-time DMET equations of motion can now be written as
\begin{equation}
i\hbar|\dot{c}\rangle=\sum_d|d\rangle X_{dc}\label{eqn:eomorb}
\end{equation}
for the embedding orbitals, where the elements $X_{dc}$ are given in Eqs. \eqref{eqn:xcorevirt}-\eqref{eqn:xbathcore},  and
\begin{equation}
i\hbar\dot{C}_M=\sum_N\langle M|\hat{H}|N\rangle C_N,\label{eqn:eomc}
\end{equation}
for the wavefunction coefficients, 
where Eq. \eqref{eqn:eomc} is obtained from Eq. \eqref{eqn:dotc} by noticing that
the matrix elements $\langle M | \hat{X} | N\rangle$ are non-zero only for intraspace rotations,
and those components of $\hat{X}$ are all defined to be zero.

It is important to note that the equations of motion for real-time DMET are similar to those derived for TD-CASSCF, which will be presented in the next section.\cite{Mir11,Sat13,Lin15} 
The main difference is that in real-time DMET, a subset of the active space orbitals, specifically the impurity orbitals, are restricted to be time-independent. This difference ensures that the definition of the impurity remains during the dynamics, which is a necessity for future extensions of the methodology to multiple impurities.
Analogous to the static DMET, 
the real-time DMET is exact for the impurity properties in the non-interacting limit, in the limit when the size of the impurity becomes
(half) the size of the full quantum system, and in the limit where there is no coupling between the impurity and the environment.
The latter exact property is not ensured by TD-CASSCF. 

\subsection{Time-Dependent CASSCF}\label{sec:rtdmet}

We conclude the theory section by introducing the working equations of motion for TD-CASSCF. The derivation of these equations have been worked out previously, and bare close resemblance to the derivation of the real-time DMET equations of motion, so here we only provide the final results.\cite{Mir11,Sat13,Lin15}

The time-dependent wavefunction in TD-CASSCF takes an analogous form to Eq. \eqref{eqn:psi_imp} where, 
\begin{equation}
|\Psi_{\textrm{CAS}}(t)\rangle=\sum_MC_M(t)|M(t)\rangle,
\end{equation}
$|M(t)\rangle$ are time-dependent determinants comprised of the active space coupled to the doubly-occupied core orbitals, and $C_M(t)$ are the time-dependent expansion coefficients.

The equations of motion for the expansion coefficients and single-particle orbitals are
\begin{equation}
i\hbar \dot{C}_M=\sum_N\langle M|\hat{H}|N\rangle C_N\label{eqn:caseomc}
\end{equation}
and
\begin{equation}
i\hbar|\dot{c}\rangle=\sum_d|d\rangle X_{dc},\label{eqn:caseomorb}
\end{equation}
respectively,
where, as in Sec. \ref{sec:rtdmet}, the single-particle Hermitian operator $\hat{X}$ governs the time-dependence of the orbitals. In comparison to real-time DMET, however, in TD-CASSCF all of the active space orbitals are time-dependent.

The non-redundant elements of the operator $\hat{X}$ are given by
\begin{eqnarray}
X_{au}=X_{ua}^*&=&\tilde{h}_{au}+\sum_{lk}\left(V_{kalu}-\frac{1}{2}V_{kaul}\right)\rho_{lk}^{\Psi_{\textrm{CAS}}}\nonumber\\
&+&\sum_v\left(2V_{vavu}-V_{vauv}\right),\label{eqn:casX1}\\
X_{an}=X_{na}^*&=&\tilde{h}_{an}+\sum_l\left(\sum_{vk}\rho_{kl}^{\Psi_{\textrm{CAS}}}\left(2V_{vavk}-V_{vkuv}\right)\right.\nonumber\\
&+&\left.\sum_{jkm}V_{jamk}\Gamma^{\Psi_{\textrm{CAS}}}_{kmlj}\right)\left[\left(\rho^{\Psi_{\textrm{CAS}}}\right)^{-1}\right]_{ln},\label{eqn:casX2}
\end{eqnarray}
and
\begin{eqnarray}
X_{nu}=X_{un}^*&=&\tilde{h}_{nu}+\sum_k\left[\left(\bar{\rho}^{\Psi_{\textrm{CAS}}}\right)^{-1}\right]_{nk}\nonumber\\
&&\left(2\sum_{v}\left(2V_{vkvu}-V_{vkuv}\right)-\sum_{jlm}V_{jlmu}\Gamma^{\Psi_{\textrm{CAS}}}_{mkjl}\right.\nonumber\\
&+&\sum_{lm}\rho^{\Psi_{\textrm{CAS}}}_{lm}\left(2V_{mklu}-V_{mkul}\right)\nonumber\\
&-&\left.\sum_{lv}\rho^{\Psi_{\textrm{CAS}}}_{kl}\left(2V_{vlvu}-V_{vluv}\right)\right),\label{eqn:casX3}
\end{eqnarray}
where  $\rho^{\Psi_{\mathrm{CAS}}}_{ij}=\langle \Psi_{\mathrm{CAS}}|E_{ji}|\Psi_{\mathrm{CAS}}\rangle$ and $\left[\bar{\rho}^{\Psi_{\mathrm{CAS}}}\right]_{yz}=2\delta_{yz}-\rho^{\Psi_{\mathrm{CAS}}}_{yz}$.
All other elements of the operator $\hat{X}$ constitute redundant orbital rotations and are thus set to zero as in Sec. \ref{sec:rtdmet}.
Mirroring the notation from the previous sections, in this section, active space orbitals are designated by, $j, k, l, m$ and $n$; doubly occupied core orbitals are designated by $u$ and $v$; unoccupied virtual orbitals are designated by $a$; the combination of all single-particle orbitals are designated by $c$ and $d$.

Equations \eqref{eqn:casX1}-\eqref{eqn:casX3} involve the inverses of $\rho^{\Psi_{\mathrm{CAS}}}$ and $\bar{\rho}^{\Psi_{\mathrm{CAS}}}$, which become numerically unstable when a subset of the single particle orbitals in the active space become fully unoccupied or occupied, respectively. To avoid this numerical instability we regularize both matrices following a similar procedure to what is done in multiconfiguration time-dependent Hartree theory.\cite{Man92,Mey90} Specifically, any eigenvalue of the matrices below a small threshold $\varepsilon$ is set to $\varepsilon$.


\section{Single impurity Anderson model}\label{sec:siam}

In this work, we will compare real-time DMET and TD-CASSCF for the simulation of the non-equilibrium dynamics of the single impurity Anderson model (SIAM).
The SIAM is a model of an interacting impurity embedded in a non-interacting environment,
and can be realized in  different physical systems, such as in quantum dots or molecules attached
to metallic leads.\cite{Kel14,Ama13,Koe16} As the simplest example of a bulk interacting quantum problem,
it provides a useful benchmark system for non-equilibrium electron dynamics in the presence of electron correlation. We emphasize, however, that both methods are applicable to more general and complex systems.

\begin{figure}
\includegraphics[scale=0.35]{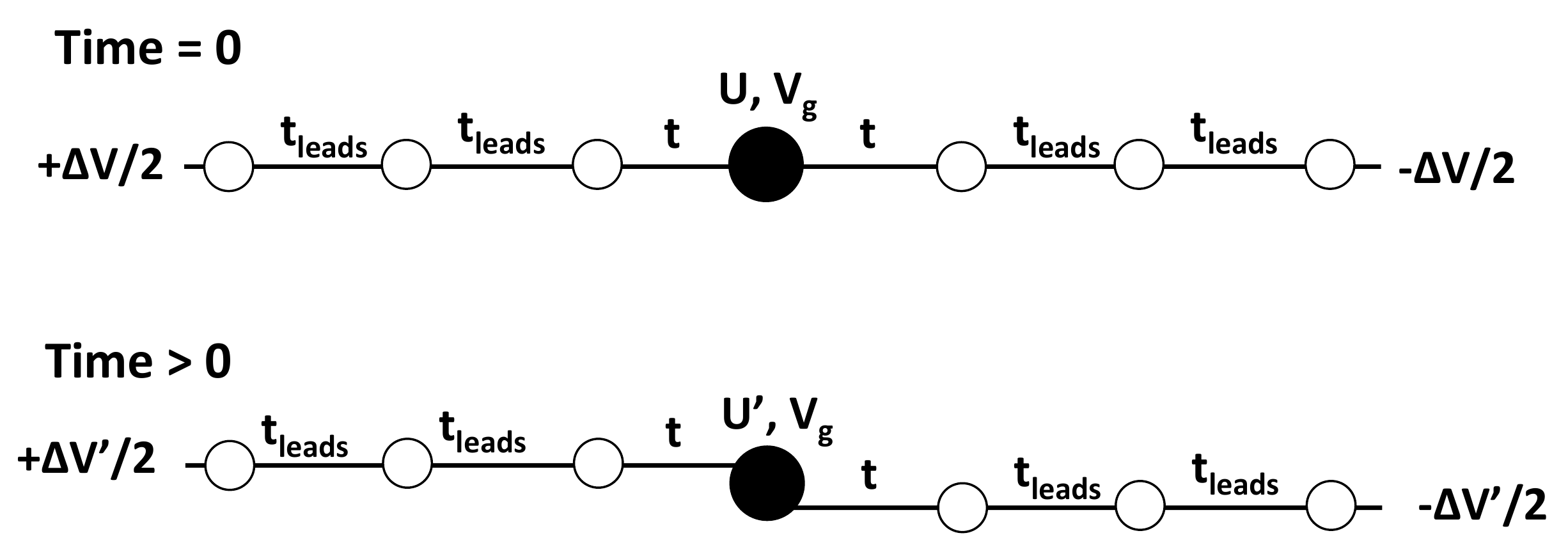}
\caption{\label{fig:siam} A pictorial representation of the quantum quenches studied in this work in the single impurity Anderson model (SIAM). The SIAM consists of a quantum dot (solid black circle) with a local Coulomb interaction, $U$, and gate-potential, $V_g$, coupled to two non-interacting leads (open circles) with hopping $t$; the lead sites are coupled with hopping $t_{\mathrm{leads}}$. In addition, a bias, $\Delta V$, can be applied across the leads. The initial state of the system at time = 0 is calculated as the ground-state of the SIAM described by a Hamiltonian defined by parameters $U$ and $\Delta V$. The subsequent dynamics at time $>$ 0 are then run using a Hamiltonian defined by parameters $U'$ and $\Delta V'$.}
\end{figure}

The SIAM consists of a single quantum dot site, where Coulomb interactions are present, coupled to two non-interacting leads, Fig. \ref{fig:siam}. In this work, we use a real-space definition of the SIAM, in which the leads have nearest neighbor hopping terms, to allow for easy comparison with real-time DMRG calculations.\cite{Alhas06} The leads are finite in size and the total system size including
the leads and interacting site is $N$.
The SIAM under a bias is then described by the Hamiltonian $\hat{H}=\hat{H}_{\mathrm{dot}}+\hat{H}_{\mathrm{leads}}+\hat{H}_{\mathrm{dot-leads}}+\hat{H}_{\mathrm{bias}}$ where
\begin{equation}
\hat{H}_{\mathrm{dot}}=V_gn_d+Un_{d\uparrow}n_{d\downarrow}
\end{equation}
describes the quantum dot in isolation. The quantum dot is located at site $d=N/2$, $V_g$ is the gate potential which controls the location of the energy level of the quantum dot, $U$ is the local Coulombic interaction, $n_d=n_{d\uparrow}+n_{d\downarrow}$, and $n_{d\sigma}=a^\dag_{d\sigma}a_{d\sigma}$. The Hamiltonian of the leads in isolation is
\begin{equation}
\hat{H}_{\mathrm{leads}}=-t_{\mathrm{leads}}\sum_{p\sigma}\left(a^\dag_{Lp\sigma}a^\dag_{Lp+1\sigma}+a^\dag_{Rp\sigma}a^\dag_{Rp+1\sigma}+h.c.\right),
\end{equation}
where $t_{\mathrm{leads}}$ is the hopping amplitude of the lead and the subscript $Lp$ ($Rp$) denotes site $p$ in the left (right) lead. The quantum dot is coupled to the two leads through the term
\begin{equation}
\hat{H}_{\mathrm{dot-leads}}=-t\sum_\sigma\left(a^\dag_{L1\sigma}a_{d\sigma}+a^\dag_{R1\sigma}a_{d\sigma}+h.c.\right),
\end{equation}
where $t$ describes the hopping amplitude between the surrounding leads and the quantum dot and the subscript $L1$ ($R1$) denotes the lead site that is closest to the quantum dot in the left (right) lead. Lastly, a bias can be applied across the SIAM, of
the form
\begin{eqnarray}
\hat{H}_{\mathrm{bias}}=\frac{\Delta V}{2}\sum_{p\sigma}\left(a^\dag_{Lp\sigma}a_{Lp\sigma}-a^\dag_{Rp\sigma}a_{Rp\sigma}\right).
\end{eqnarray}

In this work, we investigate the non-equilibrium dynamics following two types of quantum quenches, depicted in Fig. \ref{fig:siam}. In the first, the initial state for the subsequent dynamics is defined as the ground-state of the SIAM under zero bias, $\Delta V=0$; the dynamics of the initial state are then propagated using a Hamiltonian including a finite bias, $\Delta V'\ne0$, which drives a current through the quantum dot. In the second, the initial state is defined as the ground-state of the SIAM with a specific value of the local Coulomb interaction, $U$; the dynamics are then propagated using a Hamiltonian with a different interaction, $U'\ne U$.

The dynamics following the quantum quenches are characterized through several observables. Specifically, we investigate the time-dependence of the occupancy on the quantum dot, $n_{\mathrm{d}}$, and the time-dependent current through the dot, which is defined as the average of the current between the dot and the closest left lead-state and closest right lead-state, $J(t)=\left(J_{\mathrm{L}}(t)+J_{\mathrm{R}}(t)\right)/2$, where\cite{Alhas06,Wan13}
\begin{equation}
J_{\mathrm{L}}(t)=-\frac{ite}{\hbar}\sum_{\sigma}\langle\Psi_{\mathrm{imp}}(t)|a_{L1\sigma}^\dag a_{d\sigma}-a_{d}^\dag a_{L1\sigma}|\Psi_{\mathrm{imp}}(t)\rangle,\label{eqn:JL}
\end{equation}
and
\begin{equation}
J_{\mathrm{R}}(t)=-\frac{ite}{\hbar}\sum_{\sigma}\langle\Psi_{\mathrm{imp}}(t)|a_{d}^\dag a_{R1\sigma}-a_{R1\sigma}^\dag a_{d\sigma}|\Psi_{\mathrm{imp}}(t)\rangle.\label{eqn:JR}
\end{equation}
The use of the symmetrized current provides better numerical convergence to infinite system size, particularly when the left and right leads have a different number of sites.\cite{Alhas06,Wan13} The conductance, $G$, can be obtained by dividing the steady-state value of the current by the total bias applied across the leads during the dynamics, $\Delta V'$. 

\section{Calculation Details}

For calculations utilizing real-time DMET, the initial state of the system is calculated using the static DMET algorithm described in Sec. \ref{sec:statdmet} using exact diagonalization as the impurity solver. 
  The subsequent dynamics of the electronic wavefunction are propagated using the real-time DMET equations of motion, Eqs. \eqref{eqn:eomorb} and \eqref{eqn:eomc}, which are evaluated using the fourth-order Runge-Kutta method. All results are fully converged with a time-step of 0.005. 
  
  We present real-time DMET results with varying impurity size, $N_{\mathrm{imp}}$, which is kept the same between the initial static and subsequent real-time DMET calculations. The impurity always includes the quantum dot, followed by lead states in increasing distance from the quantum-dot; a left-lead state is always included prior to a right-lead state, though results are relatively insensitive to this choice. Thus, an impurity of size $N_{\mathrm{imp}}=3$ includes the quantum dot and the closest lead-state from the right and left leads, while an impurity size of $N_{\mathrm{imp}}=4$ includes the quantum dot, the two closest lead-states from the left lead, and the closest lead state from the right lead.
  
  For calculations utilizing TD-CASSCF, the initial state of the system is calculated using CASSCF as implemented in the PySCF package.\cite{Sun17}
   The subsequent dynamics of the electronic wavefunction are propagated using the TD-CASSCF equations of motion, Eqs. \eqref{eqn:caseomc} and \eqref{eqn:caseomorb}, which are evaluated using the fourth-order Runge-Kutta method. All results are fully converged with a time-step of 0.0001 and a value of the regularization parameter $\varepsilon=10^{-8}$.

We present TD-CASSCF results with varying active space size, $N_{\mathrm{CAS}}$, which is kept the same between the initial static and subsequent time-dependent CASSCF calculations. The active space is always chosen to include the HOMO through HOMO-$N_{\mathrm{CAS}}/2$ and LUMO through LUMO+$N_{\mathrm{CAS}/2}$ orbitals from the Hartree-Fock calculation used to initialize the CASSCF calculation.

As a reference, we compare results from real-time DMET and TD-CASSCF to results generated using TD-DMRG either computed
using the \textsc{ITensor} library\cite{itensor} with a bond-dimension of 300 or from previous work.\cite{Alhas06} Comparisons are also made to results generated using TDHF theory.\cite{Dir30,Fre34,Dre05}

\section{Results}

We now present our results utilizing real-time DMET to simulate the non-equilibrium electron dynamics in the SIAM following a variety of quantum quenches. In all cases, the parameter $t_{\mathrm{leads}}=1.0$ is taken as the energy scale and the parameter $t=0.4$; the parameters $V_g$, $U$, $U'$, $\Delta V$, and $\Delta V'$ are varied to define a wide range of quantum quenches.

\begin{figure}
\includegraphics[scale=0.95]{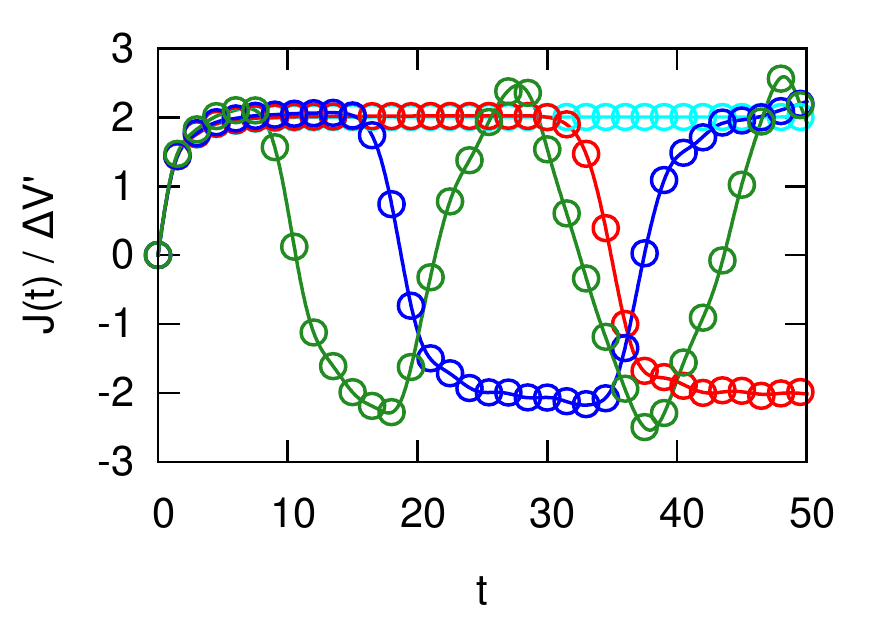}
\caption{\label{fig:U0_biasquench} The time-dependent current, $J(t)$, following a non-interacting quench in which a bias is suddenly switched on calculated exactly (open circles) and with real-time DMET (solid lines) with $N=16$ (green), $N=32$ (blue), $N=64$ (red), and $N=128$ (cyan). The parameters are $U=U'=0$, $\Delta V=0$, and $\Delta V'=-0.001$. }
\end{figure}

As mentioned in Sec. \ref{sec:theory}, real-time DMET is exact in the non-interacting limit. This is numerically verified in Fig. \ref{fig:U0_biasquench} in which we present results for a non-interacting quantum quench in which a bias is suddenly switched on to drive current through the quantum dot; the parameters are $U=U'=0$, $V_g=0$, $\Delta V=0$, and $\Delta V'=-0.001$. Fig. \ref{fig:U0_biasquench} illustrates that the current through the quantum dot evaluated using real-time DMET (open circles) exactly matches results from exact dynamics (solid lines) for a range of  total system sizes: $N=16$ (green), $N=32$ (blue), $N=64$ (red), and $N=128$ (cyan). The real-time DMET calculations use an impurity size of $N_{\mathrm{imp}}=3$;  real-time DMET is exact in the non-interacting limit regardless of impurity size.
The exact dynamics are obtained by integrating the equations of motion for the one-electron reduced density matrix of the total system, $i\hbar\dot{\rho}_{pq}=\sum_r \rho_{pr}h_{rq}-h_{pr}\rho_{rq}$.

Figure \ref{fig:U0_biasquench} also illustrates an important result regarding the non-equilibrium electron dynamics in a finite-size SIAM. Specifically, the current is seen to oscillate for small total system size. This can be attributed to a recurrence of the electron density following a reflection off of the end of the leads. The position and height of the recurrence provides a metric by which to benchmark the dynamics generated using real-time DMET when it is not exact, similar to using a Loschmidt echo. In addition, the steady-state behavior of the SIAM is defined as the plateau regime in between recurrences.

\begin{figure}
\includegraphics[scale=1.0]{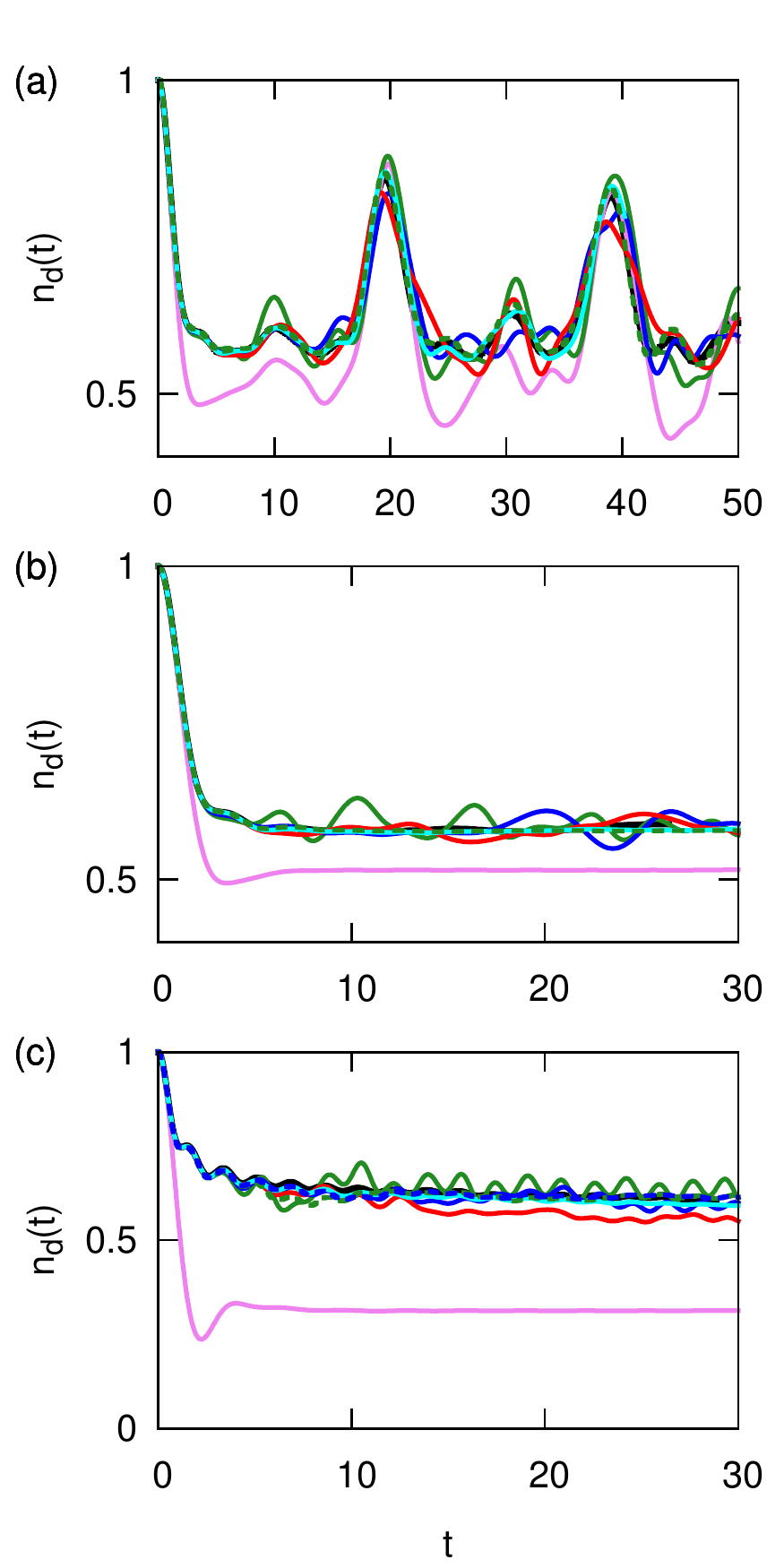}
\caption{\label{fig:int_quench} The time-dependent occupancy on the dot, $n_d(t)$, following a quantum quench in which the local Coulomb interaction on the quantum dot is suddenly switched on calculated using TD-DMRG (black)\cite{itensor}, TDHF (violet), real-time DMET with $N_{\mathrm{imp}}=3$ (green), $N_{\mathrm{imp}}=4$ (blue), $N_{\mathrm{imp}}=5$ (red), and $N_{\mathrm{imp}}=6$ (cyan), and TD-CASSCF with $N_{\mathrm{CAS}}=6$ (dashed-green) and $N_{\mathrm{CAS}}=8$ (dashed-blue) for (a) $U'=1.0$ and $N=16$, (b) $U'=1.0$ and $N=64$, (c) $U'=3.0$ and $N=64$. The remaining parameters are $U=0.0$, $V_g=0.0$, and $\Delta V=\Delta V'=0.0$.}
\end{figure}

We now turn our attention to interacting quenches in which real-time DMET is only exact in the large impurity size limit. Fig. \ref{fig:int_quench} presents the time-dependent occupancy on the quantum dot, $n_d(t)$, for a quantum quench in which the local Coulomb interaction on the quantum dot is suddenly switched on; the parameters are $U=0$ and $U'\ne0$ with $V_g=\Delta V=\Delta V'=0$.
Such a quantum quench provides a useful benchmark for real-time DMET since the initial state is a non-interacting ground-state; the initial state can thus be calculated exactly, such that the embedding orbitals are the exact embedding orbitals at time $t=0$. The subsequent dynamics provide a test solely of the accuracy of the real-time DMET equations of motion.
The figure compares results calculated using real-time DMET with an impurity size of $N_{\mathrm{imp}}=3$ (green), $N_{\mathrm{imp}}=4$ (blue), $N_{\mathrm{imp}}=5$ (red), and $N_{\mathrm{imp}}=6$ (cyan) to those that we have calculated using TD-CASSCF with an active space size of $N_{\mathrm{CAS}}=6$ (dashed-green) and $N_{\mathrm{CAS}}=8$ (dashed-blue) and with time-dependent DMRG (black),\cite{itensor} which can be taken as the exact answer. As stated in Sec. \ref{sec:statdmet}, the size of the active-space in real-time DMET is given by twice the size of the impurity, such that a real-time DMET calculation using an impurity size of $N_{\mathrm{imp}}$ should be compared to a TD-CASSCF calculation using an active-space size of $N_{\mathrm{CAS}}=2N_{\mathrm{imp}}$.
Results calculated using TDHF (violet) are also presented as a reference mean-field calculation.

Fig. \ref{fig:int_quench}(a) presents results for a small value of the Coulomb interaction, $U'=1.0$, and for a small total system size, $N=16$. 
The numerically exact TD-DMRG dynamics are characterized by a rapid decrease in the quantum dot occupancy, $n_d(t)$, followed by high-frequency, small amplitude, oscillations around a plateau value, with recurrence peaks occurring at $t\approx20$ and $t\approx40$ associated with the small total system size. Due to the severe approximations present in the method, the mean-field TDHF results overestimate the decrease in the time-dependent occupancy and do not correctly capture the high-frequency oscillations even for the relatively small value of the Coulomb interaction; TDHF, however, is able to correctly capture the position of the recurrence peaks. In comparison, real-time DMET quantitatively captures the decrease in the occupancy as well as the position and height of the recurrence peaks even for small impurity sizes, illustrating the ability of the real-time DMET method to capture the dominant time-scales in the non-equilibrium electron dynamics. The high-frequency oscillations are captured at short times, $t<20$, by real-time DMET, though the agreement becomes worse with increasing time. The worse agreement for the high-frequency oscillations can be attributed to finite-size effects associated with the impurity; since the dynamics of the embedding states are approximate, the embedding states are unable to fully damp the recurrence dynamics present within the impurity, which leads to artificial high-frequency oscillations. As seen in Fig. \ref{fig:int_quench}, and as will be illustrated further below, these artificial high-frequency oscillations disappear with increasing impurity size. 

Fig \ref{fig:int_quench}(a) also shows that TD-CASSCF actually behaves better than real-time DMET, exhibiting quantitative agreement with TD-DMRG even for a small active space. The TD-CASSCF equations of motion do not contain the constraint that the impurity orbitals must remain time-independent as is the case for real-time DMET; this increased dynamic flexibility leads to a better approximation of the time-dependent wavefunction.
This result is not too surprising in that it mirrors what is observed for static DMET; for the case of a single impurity, it is not obvious that there is a benefit to DMET over the fully variational CASSCF.
However, Fig. \ref{fig:int_quench}(a) still shows that real-time DMET with only a single impurity rapidly converges to the exact answer, which is an important and necessary result for the development of a full multi-impurity theory.

Fig. \ref{fig:int_quench}(b) presents results for the same small value of the Coulomb interaction, $U'=1.0$, but for a larger total system size, $N=64$. For this system size, recurrence peaks are no longer present in the TD-DMRG dynamics for the time-scale pictured. Instead, the time-dependent occupancy is characterized by a rapid decay to a plateau value. Once again, the TDHF results underestimate the plateau value. Both real-time DMET and TD-CASSCF show a marked improvement over TDHF, correctly capturing the time-scale for the rapid decay and the magnitude of the plateau value. However, as in Fig. \ref{fig:int_quench}(a), the real-time DMET dynamics, in comparison to TD-CASSCF, exhibit artificial high-frequency oscillations in the plateau region, $t>5$, for small impurity sizes. The magnitude of these oscillations can be clearly seen to diminish with increasing impurity size, and the $N_{\mathrm{imp}}=6$ results exhibit no oscillations whatsoever.

\begin{figure}
\includegraphics[scale=1.0]{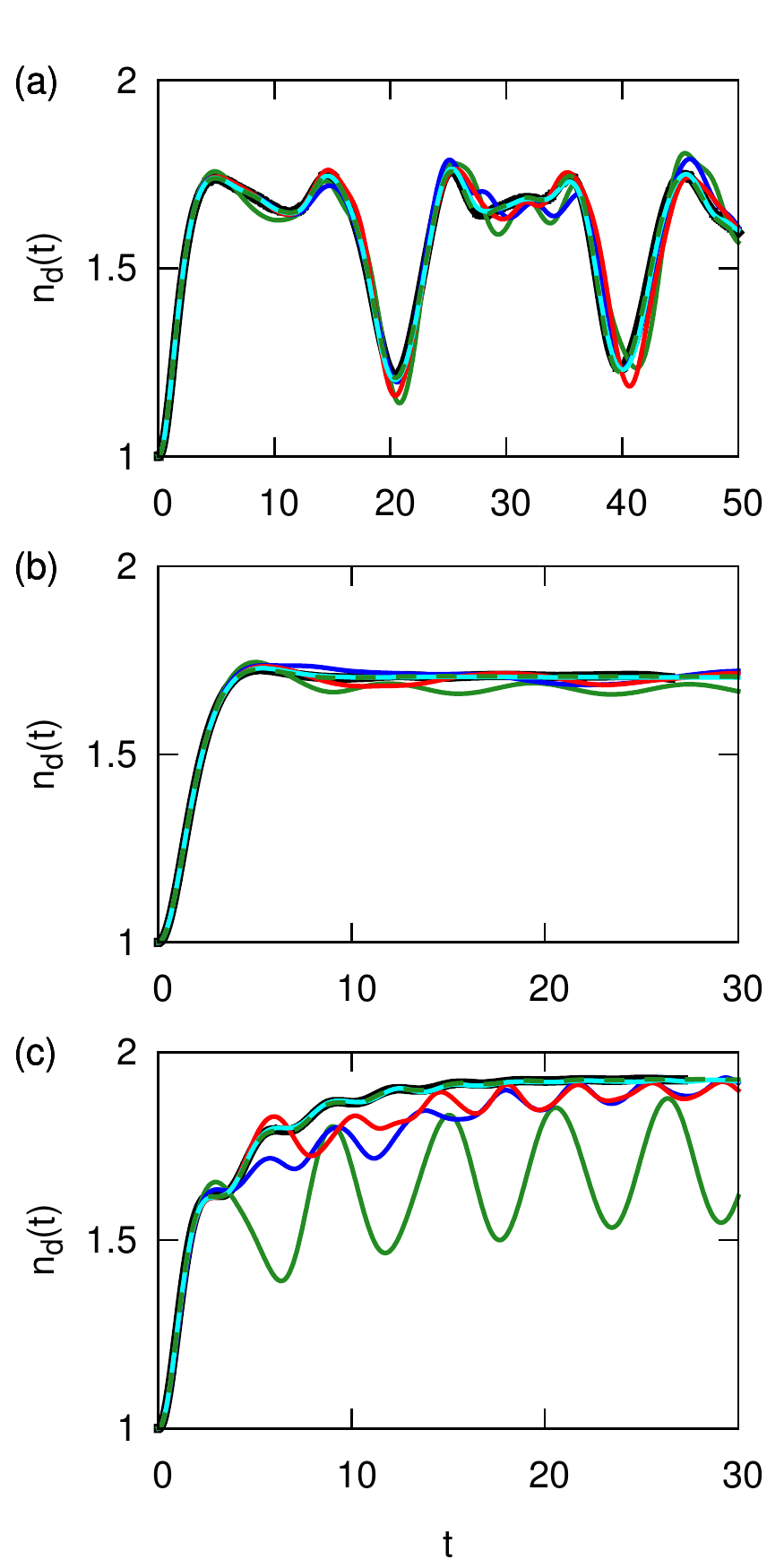}
\caption{\label{fig:nonint_quench} The time-dependent occupancy on the dot, $n_d(t)$, following a quantum quench in which the local Coulomb interaction on the quantum dot is suddenly switched off calculated using TD-DMRG (black)\cite{itensor}, real-time DMET with $N_{\mathrm{imp}}=3$ (green), $N_{\mathrm{imp}}=4$ (blue), $N_{\mathrm{imp}}=5$ (red), and $N_{\mathrm{imp}}=6$ (cyan), and TD-CASSCF with $N_{\mathrm{CAS}}=6$ (dashed-green) for (a) $U=1.0$ and $N=16$, (b) $U=1.0$ and $N=64$, and (d) $U=3.0$ and $N=64$. The remaining parameters are $U'=0.0$, $V_g=-U/2$, and $\Delta V=\Delta V'=0.0$.}
\end{figure}

Fig. \ref{fig:int_quench}(c) present results for a larger value of the Coulomb interaction, $U'=3.0$, and for the larger system size, $N=64$. This provides a more stringent test for real-time DMET as a larger Coulomb interaction leads to stronger correlation between the the impurity and the surrounding environment. 
The TD-DMRG results show similar behavior to the $U'=1.0$ results; the results are characterized by a rapid decay followed by a plateau. In comparison to the $U'=1.0$ results, though, the dynamics also exhibit high-frequency oscillations throughout the entire trajectory. 
Expectantly, the TDHF results are worse than those obtained in Figs. \ref{fig:int_quench}(a) and (b); the mean-field approximation becomes worse for higher values of the Coulomb interaction due to the presence of strong correlation.
In comparison, both real-time DMET and TD-CASSCF are again able to correctly capture the rapid decay of the occupancy and the value of the plateau. Though as before, real-time DMET exhibits artificial high-frequency oscillations for small impurity sizes, which disappear as the size of the impurity is increased.

Taken together, the results in Fig. \ref{fig:int_quench} clearly illustrate that real-time DMET with very small
impurities provides a qualitatively correct picture of the non-equilibrium dynamics,
and that accuracy increases rapidly with the impurity size.
In fact, the $N_{\mathrm{imp}}=6$ results are in almost quantitative agreement with the TD-DMRG results for all of the presented cases. This illustrates the computational benefits of embedding, since in comparison to the full exact calculation, which would involve $N$ correlated sites, the real-time DMET method is able to provide the same description using only $2N_{\mathrm{imp}}$  correlated 
orbitals ($N_{\mathrm{imp}}$ impurity and $N_{\mathrm{imp}}$ embedding bath orbitals).


In comparison to the calculations performed in Fig. \ref{fig:int_quench}, most realistic simulations will not involve an exactly 
computed initial state. As such, Fig. \ref{fig:nonint_quench} presents the time-dependent occupancy on the quantum dot, $n_d(t)$, for a quantum quench in which the local Coulomb interaction on the quantum dot is suddenly switched off; the parameters are $U\ne0$ and $U'=0$ with $V_g= U/2$ and $\Delta V=\Delta V'=0$. 
This quantum quench presents a situation in which the initial state obtained using
static DMET (or CASSCF) is no longer exact; the subsequent dynamics are thus also not exact even though the dynamics are generated using a non-interacting Hamiltonian.
The figure again compares results calculated using real-time DMET with an impurity size of $N_{\mathrm{imp}}=3$ (green), $N_{\mathrm{imp}}=4$ (blue), $N_{\mathrm{imp}}=5$ (red), and $N_{\mathrm{imp}}=6$ (cyan) to those we have calculated using TD-CASSCF with an active space size of $N_{\mathrm{CAS}}=6$ (dashed-green) and TD-DMRG (black),\cite{itensor} which can be taken as the exact answer. Results using TDHF are no longer shown as even the initial static Hartree-Fock guess is inaccurate.

The real-time DMET results in Fig. \ref{fig:nonint_quench} show very similar behavior to that seen in Fig. \ref{fig:int_quench}. Fig. \ref{fig:nonint_quench}(a) presents the results for a small total system size, $N=16$, and a small value of the Coulomb interaction, $U=1.0$. Again, the real-time DMET results are able to capture the position and amplitude of the recurrence peaks at $t\approx20$ and $t\approx40$ regardless of impurity size, while the small impurity size results show artificial high-frequency oscillations. Similarly, Fig. \ref{fig:nonint_quench}(b) illustrates that the real-time DMET results correctly capture the time-scale of the rapid increase in occupancy on the dot and the correct height of the plateau region for a larger total system size, $N=64$, regardless of impurity size. The small impurity size results exhibit oscillations, which disappear with increasing impurity size. Lastly, Figs. \ref{fig:nonint_quench}(d) present results for $U=3.0$ and a large, $N=64$, total system size, respectively. As observed in Fig. \ref{fig:int_quench}, the agreement between real-time DMET and TD-DMRG is not as good at small impurity size for the larger value of $U$. However, the real-time DMET results are clearly seen to converge to the TD-DMRG reference results for $N_{\mathrm{imp}}=6$. As was seen in Fig. \ref{fig:int_quench}, Fig. \ref{fig:nonint_quench} shows that TD-CASSCF converges faster than real-time DMET with respect to size of the active space; TD-CASSCF exhibits quantitative agreement, even for the larger value of $U=3.0$, with TD-DMRG for the small active space size of $N_{\mathrm{CAS}}=6$.

\begin{figure}
\includegraphics[scale=1.0]{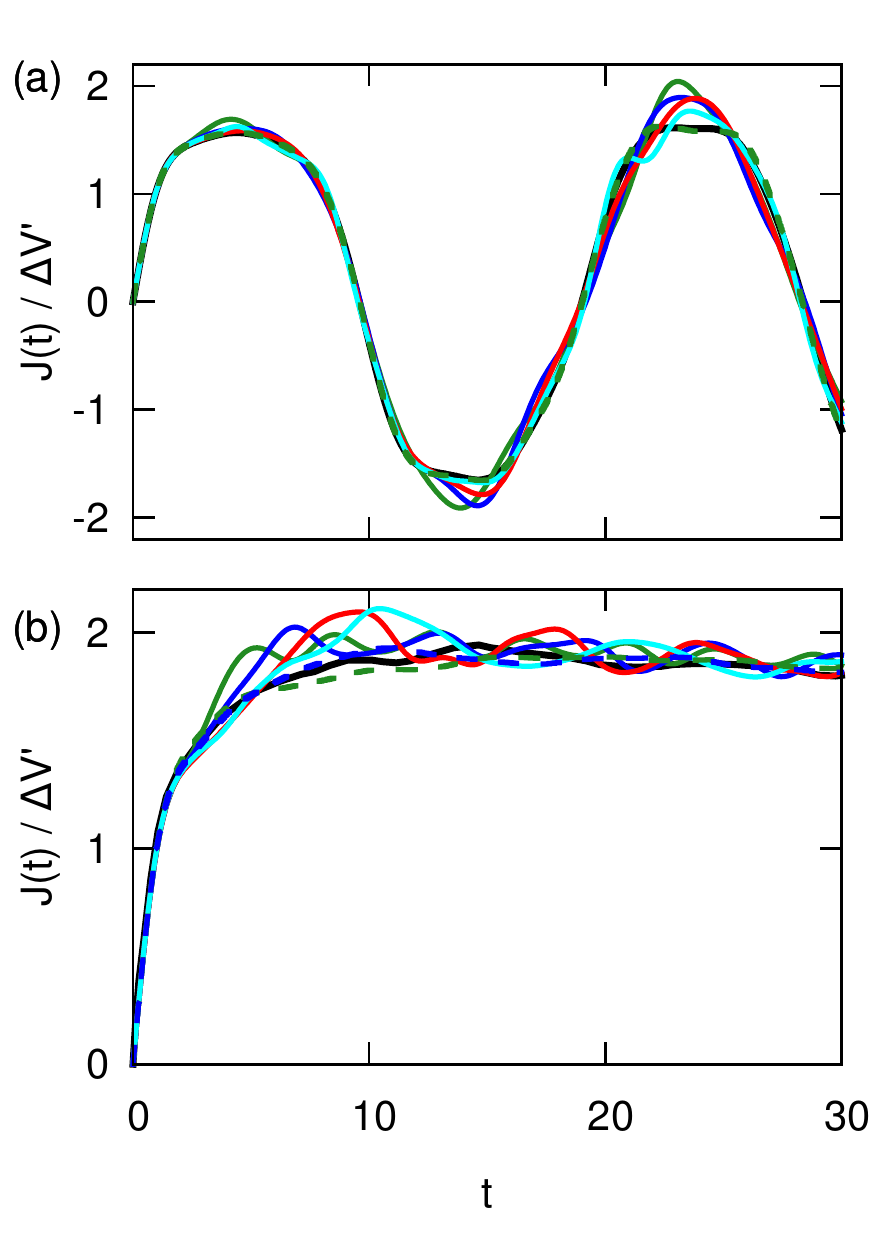}
\caption{\label{fig:biasquench} The time-dependent current, $J(t)$, through the quantum dot for an interacting quench in which a bias is suddenly switched on calculated with TD-DMRG (black)\cite{Alhas06} and real-time DMET with $N_{\mathrm{imp}}=3$ (green), $N_{\mathrm{imp}}=4$ (blue), $N_{\mathrm{imp}}=5$ (red), and $N_{\mathrm{imp}}=6$ (cyan) for (a) $N=16$ and (b) $N=128$. The parameters are $U=U'=1.0$, $V_g=-0.5$, $\Delta V=0$, and $\Delta V'=-0.005$.}
\end{figure}

We conclude this section by investigating an interacting quantum quench in which a bias is suddenly switched on to drive current through the quantum dot.
Fig. \ref{fig:biasquench} presents the time-dependent current through the quantum dot calculated using TD-DMRG from Ref. \citenum{Alhas06} (black),  real-time DMET with an impurity size of $N_{\mathrm{imp}}=3$ (green), $N_{\mathrm{imp}}=4$ (blue), $N_{\mathrm{imp}}=5$ (red), and $N_{\mathrm{imp}}=6$ (cyan), and TD-CASSCF with an active space size of $N_{\mathrm{CAS}}=6$ (dashed-green) and $N_{\mathrm{CAS}}=8$ (dashed-blue); the parameters are $U=U'=1.0$, $V_g=U/2$, $\Delta V=0$ and $\Delta V'=-0.005$. Fig. \ref{fig:biasquench}(a) presents results for a small total system size of $N=16$. As seen in the non-interacting case, Fig. \ref{fig:U0_biasquench}, the small total system size leads to oscillations of the current. The frequency of these oscillations are captured by real-time DMET regardless of impurity size; the amplitude of the initial oscillation is captured by all impurity sizes, while the amplitude of subsequent oscillations are accurately captured only by the larger impurity size calculations. TD-CASSCF quantitatively matches TD-DMRG for all times even small active space size. Such a result corroborates the behavior observed in the previous quantum quenches, in which it is necessary to push real-time DMET to larger impurity size to correctly capture the longer-time dynamics, while TD-CASSCF converges to the exact answer even for small active space sizes.

Fig. \ref{fig:biasquench}(b) presents results for a larger total system size of $N=128$. The TD-DMRG results exhibit a rapid increase of the current, followed by a plateau region characterized by low amplitude oscillations;\cite{Alhas06} the oscillations in the plateau region have been extensively discussed in the literature\cite{Alhas06,Har15,Hei09,Sch00,Sch08} and can be attributed to the level spacing within the leads associated with the finite-size of the total system.
The real-time DMET and TD-CASSCF results are able to correctly capture both the time-scale for the increase of the current as well as the plateau value of the current. 
It is important to emphasize that the capacity of real-time DMET to correctly capture the plateau value of the current 
under these conditions is a non-trivial result. In principle, the value of the gate-voltage, $V_g=-U/2$, should put the system in the conductance ``valley", such that no current should be observed. However, the Kondo effect, which arises from the coupling between the localized spin on the quantum dot and the conducting electrons in the leads, yields a non-zero value of the current.\cite{Gla88,Pus06,Fer99,Dav02} The ability of real-time DMET to simulate this many-body effect illustrates the power of the method to treat dynamics
in the presence of strong correlations between the impurity and its environment.

\begin{figure*}
\includegraphics[scale=1.0]{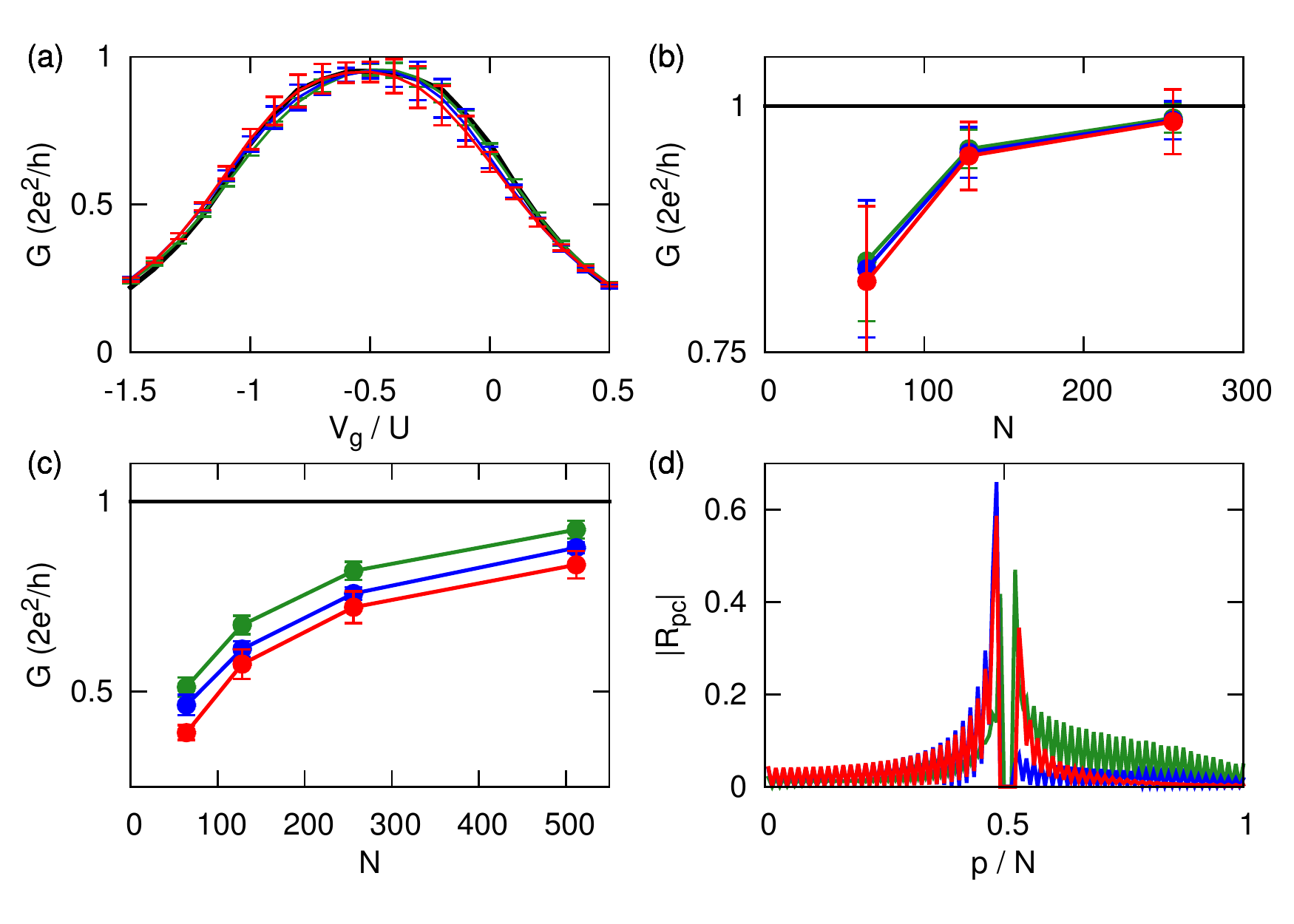}
\caption{\label{fig:conductance} 
The conductance for an interacting quench in which a bias is suddenly switched on calculated with real-time DMET with $N_{\mathrm{imp}}=3$ (green), $N_{\mathrm{imp}}=4$ (blue), and $N_{\mathrm{imp}}=5$ (red). The conductance is calculated as a function of (a) the gate potential $V_g$ for $U=U'=1.0$, (b) the total system size, $N$, for $U=U'=1.0$ and (c) the total system size, $N$, for $U=U'=4.0$. The parameters are $\Delta V=0$, and $\Delta V'=-0.005$. In part (a), the black line corresponds to the conductance calculated using TD-DMRG, while in parts (b) and (c), the black lines correspond to the ideal limit of the conductance, $2e^2/h$. Part (d) presents the magnitude of the coefficients of an embedding bath orbital corresponding to part (c) for $N=128$, $|R_{pc}|$ (Eq. \eqref{eqn:rotmat}), as a function of site index normalized by the total system size, $p/N$; the quantum dot is located at $p/N=0.5$ and the color scheme matches parts (a)-(c).}
\end{figure*}

To further emphasize this point, Fig. \ref{fig:conductance}(a), plots the conductance for the interacting quantum quench corresponding to Fig. \ref{fig:biasquench}(b) as a function of the gate potential $V_g$. As was done for the time-dependent DMRG calculations,\cite{Alhas06} the conductance is calculated as the average over the oscillations in the plateau region of the time-dependent current; the error bars report on the error associated with this average. Fig. \ref{fig:conductance}(a) compares the conductance calculated using time-dependent DMRG\cite{Alhas06} (black) and real-time DMET with an impurity size of $N_{\mathrm{imp}}=3$ (green), $N_{\mathrm{imp}}=4$ (blue) and $N_{\mathrm{imp}}=5$ (red). The figure illustrates that the conductance is correctly calculated using real-time DMET for all values of $V_g$ even for the small impurity size of $N_{\mathrm{imp}}=3$. At $V_g = -U/2$ the exact conductance is $2e^2/h$;
the real-time DMET reproduces this result just as accurately as the time-dependent DMRG. 

To achieve the exact conductance it is necessary to push to larger total system sizes.
Fig. \ref{fig:conductance}(b) illustrates one of the main benefits of the real-time DMET method,  the ability to treat significantly larger system sizes compared to non-embedding methods. Fig. \ref{fig:conductance}(b) presents the conductance as a function of total system size obtained from real-time DMET with an impurity size of $N_{\mathrm{imp}}=3$ (green), $N_{\mathrm{imp}}=4$ (blue) and $N_{\mathrm{imp}}=5$ (red). 
The conductance is obtained for the same interacting quantum quench as in Figs. \ref{fig:conductance}(a) and \ref{fig:biasquench}(b) with $V_g=-U/2$.
Fig. \ref{fig:conductance}(b) illustrates that by pushing to $N=256$, the ideal limit of the conductance (black line) is recovered; results for smaller system sizes are also pictured to show the convergence with respect to total system size.

Fig. \ref{fig:conductance}(c) illustrates that this behavior is not limited to small values of the Coulomb interaction.
Fig. \ref{fig:conductance}(c) presents the conductance as a function of total system size for a large value of the Coulomb interaction $U=U'=4.0$; the color scheme is the same as above and the remaining parameters are $V_g=-2.0$, $\Delta V=0$, and $\Delta V'=-0.005$.
The large value of the Coulomb interaction leads to finite size effects of the total system size due to the large size of the Kondo cloud.
Fig. \ref{fig:conductance}(c) shows that the value of the conductance at $N=128$, which was the size of the system used in Fig. \ref{fig:conductance}(a) and corresponds to the largest
system  studied previously with time-dependent DMRG,\cite{Alhas06} is significantly below the ideal limit of $2e^2/h$. However, 
with DMET, we can easily increase the system size. 
The remaining points in Fig. \ref{fig:conductance}(c) show that the conductance approaches the ideal limit as we increase
the total system size from 128 to 512 sites.

Lastly, Fig. \ref{fig:conductance}(d) highlights why real-time DMET is able to capture the Kondo effect. Fig. \ref{fig:conductance}(d) presents the magnitude of the coefficients of an embedding bath orbital at a single point in time as a function of site index for the interacting quantum quench presented in Fig. \ref{fig:conductance}(c); the total system size corresponds to $N=128$. The figure shows that the bath orbital is delocalized across the leads, allowing for a proper treatment of the delocalized Kondo cloud.

\section{Conclusions}

In this work, we present an extension of the density matrix embedding theory (DMET) to simulate real-time non-equilibrium electron dynamics. 
Like in the static case, the real-time DMET method partitions the full system into an impurity and an environment. The environment is efficiently represented by a quantum-bath of the same size as the impurity. The dynamics of the embedding problem, which consists of the impurity coupled to the quantum bath, is obtained through the time-dependent variational principle, in which we introduce the constraint that the impurity single-particle orbitals are time independent; such a constraint maintains the definition of the impurity during the dynamics and ensures the highest fidelity representation of the Hilbert space is retained for the impurity region.


The accuracy of the real-time DMET method has been benchmarked through comparisons with time-dependent Hartree-Fock (TDHF) theory, time-dependent complete active space self consistent field (TD-CASSCF) theory, and time-dependent density
matrix renormalization group (DMRG)\cite{itensor,Alhas06}
for a variety of quantum quenches in the single impurity Anderson model (SIAM). 
We have shown that real-time DMET shows a marked improvement over the mean-field TDHF, correctly capturing the rapid change in the occupancy of the quantum dot following the sudden switching on or off of the local Coulomb interaction on the dot regardless of the size of the impurity.
Furthermore, real-time DMET rapidly converges with respect to impurity size, to quantitatively capture recurrence peaks for small total system size, or the correct plateau behavior for large total system size, for a range of values of the Coulomb interaction. However, real-time DMET exhibits artificial oscillations in the dynamics for small values of the impurity size, which are not observed in the TD-CASSCF dynamics. These oscillations are attributed to the finite-size error of the size of the impurity and the lack of additional dynamical flexibility observed in the TD-CASSF equations of motion.

Additionally, we illustrate that real-time DMET can describe the non-trivial Kondo behavior during an interacting quantum quench in which a bias is suddenly switched on across the leads in the SIAM. Real-time DMET, in agreement with time-dependent DMRG, exhibits a value of the conductance near the ideal limit, for small values of the Coulomb interaction, and small total system sizes. However,
in comparison to time-dependent DMRG, real-time DMET can also simulate the significantly larger system sizes necessary to 
recover the ideal limit of the conductance, even for large values of the Coulomb interaction.

Taken together, the results presented in this work illustrate the capability of the real-time DMET method for simulating non-equilibrium dynamics in which strong correlation plays an important role. In addition, the methodology provides a useful starting point for future extensions, as have been carried out for static DMET, such as the use of multiple impurities. Other extensions can include finite temperature effects or the use of other wavefunction ansatzes, such as a coupled-cluster or matrix product sate wavefunction, as the starting point for the equations of motion of the embedding states; utilizing a more compact wavefunction  would allow for the simulation of larger impurity sizes than those treated in this study.

\begin{acknowledgments}

This work was support by the US National Science Foundation, through CHE-1265277. We thank Miles Stoudenmire for help with \textsc{ITensor}.

\end{acknowledgments}


\appendix{

\section{$v$-representability issues of a self-consistent real-time DMET}\label{app:vrep}

As mentioned in Sec. \ref{sec:rtdmet}, other real-time extensions of DMET are, in principle, conceivable. In this appendix, we introduce a real-time formulation that is more closely analogous to the static DMET, in which the equations of motion of the embedding orbitals are derived from a reference mean-field dynamics for the full quantum system, which is propagated in tandem with the correlated dynamics of the embedding system. A self-consistency condition in terms of the impurity density can then be introduced between the reference mean-field dynamics and the correlated dynamics of the embedding problem, enforced via a time-dependent correlation potential. One advantage of this picture is that it yields a strong global consistency condition when there are multiple impurities. However, we have found that $v$-representability problems almost always occur after sufficiently long-time propagations.

The derivation of this real-time extension of DMET entails deriving the equations of motion for (i) the mean-field reduced density matrix of the full quantum system, (ii) the correlated CAS-like DMET wavefunction in the embedding problem, and (iii) the embedding orbitals. The explicit time-dependence of all terms are suppressed for clarity.

The time-dependence of the mean-field reduced density matrix is given by
\begin{eqnarray}
i\hbar\dot{\rho}^{\Phi}_{pq}=\sum_r\rho^{\Phi}_{pr}h''_{rq}-h''_{pr}\rho^{\Phi}_{rq},\label{eqn:mfeom}
\end{eqnarray}
where $\rho^{\Phi}$ is the one-particle density matrix initially obtained from the static DMET calculation and analogous to static DMET, $\hat{h}''$ is a single-particle Hamiltonian of the form
\begin{equation}
\hat{h}''=\hat{h}+\hat{u}^{\textrm{RT}}.\label{eqn:h''}
\end{equation}
The time-dependent correlation potential 
\begin{equation}
\hat{u}^{\textrm{RT}}=\sum_{p\in A}u_{pp}E_{pp} 
\end{equation}
is distinguished from the correlation potential used in the static DMET calculation, and the elements $u_{pp}^{\textrm{RT}}$ are obtained through the self-consistency condition described below.

The equations of motion for the correlated CAS-like DMET wavefunction, Eq. \eqref{eqn:psi_imp}, are derived from the time-dependent Schr\"{o}dinger equation such that,
\begin{eqnarray}
\hat{H}|\Psi_{\textrm{imp}}\rangle&=&i\hbar|\dot{\Psi}_{\textrm{imp}}\rangle\\
&=&i\hbar\sum_m\dot{C}_m|m\rangle+C_m|\dot{m}\rangle\\
&=&\sum_mi\hbar\dot{C}_m|m\rangle+C_m\hat{X}|m\rangle,
\end{eqnarray}
which yields
\begin{equation}
i\hbar \dot{C}_n=\sum_m\langle n|\hat{H}-\hat{X}|m\rangle C_m.
\end{equation}
We have once again introduced a one electron operator $\hat{X}$ which governs the time-dependence of the embedding orbitals, such that the equation of motion of the embedding orbitals are
\begin{equation}
i\hbar|\dot{c}\rangle=\sum_d|d\rangle X_{dc}.\label{eqn:Xdef}
\end{equation}

Here, as in Sec. \ref{sec:rtdmet}, the impurity orbitals are restricted to be time-independent, such that $X_{ic}=X^*_{ci}=0$. The remaining elements of $\hat{X}$ are derived using the condition that at all points in time, the embedding orbitals are eigenfunctions of the environment block of the mean-field one-particle density matrix, Eq. \eqref{eqn:rhomf}, such that
\begin{equation}
\hat{\rho}_{\textrm{env}}^{\Phi}|c\rangle=\lambda_c|c\rangle.\label{eqn:eval}
\end{equation}
The elements of $\hat{X}$ are obtained by differentiating Eq. \eqref{eqn:eval} with respect to time and using Eqs. \eqref{eqn:mfeom} and \eqref{eqn:Xdef}, yielding
\begin{equation}
X_{cd}=h_{cd}-\frac{1}{\lambda_c-\lambda_d}\langle c|\hat{h}_{\textrm{c}}''^{\dag}\hat{\rho}_{\textrm{c}}^{\Phi}-\hat{\rho}_{\textrm{c}}^{\Phi\dag}\hat{h}_{\textrm{c}}''|d\rangle,
\end{equation}
for $c\neq d$. The coupling block of the reduced density matrix, $\rho_{\textrm{c}}^{\Phi}$, is given in Eq. \eqref{eqn:rhomf}, and the coupling block of the single-particle Hamiltonian, $h_{\textrm{c}}''$, is defined analogously. The diagonal elements, $X_{cc}$, are arbitrary and can be set to zero.

The elements of the time-dependent correlation potential, $\hat{u}^{\textrm{RT}}$, are determined through a self-consistent procedure, such that, within the impurity, the mean-field density and the correlated density obtained from $|\Psi_{\mathrm{imp}}\rangle$ are equivalent at all times. This condition is achieved by minimizing the difference between the second time derivative of the two densities on the impurity; the first time-derivative of the mean-field density is independent of the correlation potential. 
However, this condition leads to $v$-representability problems, in which the second time derivative of the mean-field density becomes independent of the correlation potential.
The second time-derivative of the mean-field density on the impurity is given by
\begin{eqnarray}
\ddot{\rho}^{\Phi}_{ii}&=&-\frac{1}{\hbar^2}\sum_r\left[\left(u_{rr}-u_{ii}\right)\left(h_{ir}\rho_{ri}^{\Phi}+\rho_{ir}^{\Phi}h_{ri}\right)+\right.\nonumber\\
&&\left.\sum_q\left(h_{ir}h_{rq}\rho_{qi}^{\Phi}-h_{ir}\rho_{rq}^{\Phi}h_{qi}\right.\right.\nonumber\\
&&\left.\left.-h_{iq}\rho_{qr}^{\Phi}h_{ri}+\rho_{iq}^{\Phi}h_{qr}h_{ri}\right)\right].\label{eqn:d2dt2rhomf}
\end{eqnarray}
Taking the derivative of Eq. \eqref{eqn:d2dt2rhomf} with respect to an element of the correlation potential yields
\begin{eqnarray}\label{eqn:ddu}
\frac{\partial \ddot{\rho}^{\Phi}_{ii}}{\partial u_{rr}}=-\frac{1}{\hbar^2}\sum_r\left(h_{ir}\rho_{ri}^{\Phi}+\rho_{ir}^{\Phi}h_{ri}\right),
\end{eqnarray}
which can clearly be seen to go to zero for specific values $\rho^{\Phi}$. Numerically it is observed that these special values of $\rho^{\Phi}$ are almost always obtained for sufficiently long dynamics indicating that this formulation of real-time DMET has a $v$-representability problem.

Eq. \eqref{eqn:ddu} is analogous to the force equation used in the Runge-Gross time-dependent density functional theory derivation.\cite{Run84} In that case, however, the right-hand side can be written as $-\rho\nabla u$ under the assumption that $\rho_{r,r+\delta r}=\rho_{r,r}$ for infinitesimal $\delta r$ due to the continuity requirements on the single-particle density matrix, and thus, cannot vanish except when $\rho=0$, ensuring $v$-representability. The more severe condition encountered in the lattice formulation is due to the lack of continuity requirement on the density matrix as has been observed previously.\cite{Li08}

}

\bibliographystyle{apsrev}
\bibliography{rtdmet_biblio}

\end{document}